\documentclass[12pt,a4paper]{article}
\usepackage{fullpage}
\usepackage{amsbsy}
\usepackage{amsfonts}
\usepackage{amssymb}
\begin{document}

\begin{flushright}
\begin{tabular}{l}
hep-ph/0202152
\\
16 February 2002
\end{tabular}
\end{flushright}
\vspace{1cm}
\begin{center}
\large\bfseries
$\mathbf{\nu_{\mathit{e}}\to\nu_{\mathit{s}}}$ oscillations with large neutrino mass in NuTeV?
\\[0.5cm]
\normalsize\normalfont
Carlo Giunti
\\
\small\itshape
INFN, Sezione di Torino,
\\
\small\itshape
and
\\
\small\itshape
Dipartimento di Fisica Teorica,
Universit\`a di Torino,
\\
\small\itshape
Via P. Giuria 1, I--10125 Torino, Italy
\\[0.5cm]
\normalsize\normalfont
Marco Laveder
\\
\small\itshape
Dipartimento di Fisica ``G. Galilei'', Universit\`a di Padova,
\\
\small\itshape
and
\\
\small\itshape
INFN, Sezione di Padova,
\\
\small\itshape
Via F. Marzolo 8, I--35131 Padova, Italy
\end{center}
\begin{abstract}
We propose an explanation of the NuTeV anomaly
in terms of oscillations of electron neutrinos
into sterile neutrinos.
We derive an average transition probability
$
P_{\nu_e\to\nu_s}
\simeq
0.21 \pm 0.07
$,
which is compatible with other neutrino data
if the mass-squared difference
that drives the oscillations is
$\Delta{m}^2 \sim 10 - 100 \, \mathrm{eV}^2$.
\end{abstract}

The NuTeV collaboration measured recently \cite{Zeller:2001hh} a value of
the electroweak parameter
$\sin^2\theta_W$
higher than the standard model prediction
obtained from a fit of other electroweak data
by
\begin{equation}
\sin^2\theta_W^{\mathrm{NuTeV}}
-
\sin^2\theta_W
=
\left( 5.0 \pm 1.6 \right) \times 10^{-3}
\,.
\label{001}
\end{equation}
In this paper we propose an explanation of this discrepancy
in terms of $\nu_e\to\nu_s$ oscillations\footnote{
Other possible explanations of the NuTeV anomaly have been
discussed in Ref.~\cite{Davidson:2001ji}.
},
where $\nu_s$ is a sterile neutrino that does not participate to weak interactions
(see \cite{BGG-review-98,Gonzalez-Garcia:2002dz}).
Our analysis is approximate and based on the limited information
contained in the paper of the NuTeV collaboration \cite{Zeller:2001hh}.
A precise analysis of NuTeV data in terms of
neutrino oscillations can be done only by the NuTeV collaboration.

The NuTeV collaboration measured the ratio
of neutral current to charged current cross sections
for neutrino scattering with isoscalar targets of $u$ and $d$ quarks
\begin{equation}
R^{\nu}
=
\frac{ \sigma(\nu_\mu+N\to\nu+X) }{ \sigma(\nu_\mu+N\to\mu^-+X) }
\,,
\label{002}
\end{equation}
and the corresponding ratio
$R^{\bar\nu}$
for antineutrinos.
The Paschos-Wolfenstein relation \cite{Paschos:1973kj} allows to derive the value of
$\sin^2\theta_W$
from 
$R^{\nu}$ and $R^{\bar\nu}$:
\begin{equation}
\sin^2\theta_W
=
\frac{1}{2}
-
\frac{ R^{\nu} - r \, R^{\bar\nu} }{ 1 - r }
\,,
\label{003}
\end{equation}
where
\begin{equation}
r
\equiv
\frac{ \sigma(\bar\nu_\mu+N\to\mu^++X) }{ \sigma(\nu_\mu+N\to\mu^-+X) }
\simeq
\frac{1}{2}
\,.
\label{004}
\end{equation}

The NuTeV collaboration distinguishes CC and NC $\nu_\mu$ interactions
on the basis of the length of the track in the detector:
long tracks correspond to CC $\nu_\mu$ interactions
and
short tracks correspond to NC $\nu_\mu$ interactions,
plus a contribution from CC and NC $\nu_e$
interactions.
The number $N_{\nu_e}^{\mathrm{MC}}$ of $\nu_e$
interactions calculated through a Monte Carlo
is subtracted to the number $N^{\nu}_{\mathrm{S}}$ of short tracks
in order to obtain the number of
NC $\nu_\mu$ interactions
used to evaluate $R^{\nu}$:
\begin{equation}
R^{\nu}_{\mathrm{NuTeV}}
=
\frac{ N^{\nu}_{\mathrm{S}} - N_{\nu_e}^{\mathrm{MC}} }{ N^{\nu}_{\mathrm{L}} }
\,,
\label{005}
\end{equation}
where $N^{\nu}_{\mathrm{L}}=N_{\nu_\mu}^{\mathrm{CC}}$ is the number of long tracks.
If $\nu_e\to\nu_s$ oscillations deplete the number of
$\nu_e$
interactions with respect to the calculated one,
the number $N_{\nu_e}^{\mathrm{MC}}$ subtracted
in the numerator of Eq.~(\ref{005})
is too high, leading to a smaller value of $R^{\nu}_{\mathrm{NuTeV}}$
with respect to the true one
and,
from the Paschos-Wolfenstein relation (\ref{003}),
a value of $\sin^2\theta_W^{\mathrm{NuTeV}}$
larger than the true
$\sin^2\theta_W$.

In the presence of $\nu_e\to\nu_s$ transitions,
the true number of $\nu_e$
interactions is
\begin{equation}
N_{\nu_e}
=
N_{\nu_e}^{\mathrm{MC}}
\,
P_{\nu_e\to\nu_e}
\,,
\label{006}
\end{equation}
where
\begin{equation}
P_{\nu_e\to\nu_e}
=
1
-
P_{\nu_e\to\nu_s}
\,.
\label{0061}
\end{equation}
is the electron neutrino survival probability averaged over
the energy spectrum of the NuTeV beam.
Writing
\begin{equation}
N^{\nu}_{\mathrm{S}}
=
N_{\nu_\mu}^{\mathrm{NC}}
+
N_{\nu_e}
\,,
\label{007}
\end{equation}
from Eq.~(\ref{005})
the ratio
$R^{\nu}_{\mathrm{NuTeV}}$
reported by the NuTeV collaboration is given by
\begin{equation}
R^{\nu}_{\mathrm{NuTeV}}
=
\frac{ N_{\nu_\mu}^{\mathrm{NC}} }{ N_{\nu_\mu}^{\mathrm{CC}} }
-
\frac{ N_{\nu_e}^{\mathrm{MC}} }{ N_{\nu_\mu}^{\mathrm{CC}} }
\left( 1 - P_{\nu_e\to\nu_e} \right)
=
R^{\nu}
-
\frac{ N_{\nu_e}^{\mathrm{MC}} }{ N_{\nu_\mu}^{\mathrm{CC}} }
\left( 1 - P_{\nu_e\to\nu_e} \right)
\,,
\label{008}
\end{equation}
where
$R^{\nu}$
is the true value of the ratio (\ref{002}).
A similar relation holds for the antineutrino ratio:
\begin{equation}
R^{\bar\nu}_{\mathrm{NuTeV}}
=
R^{\bar\nu}
-
\frac{ N_{\bar\nu_e}^{\mathrm{MC}} }{ N_{\bar\nu_\mu}^{\mathrm{CC}} }
\left( 1 - P_{\nu_e\to\nu_e} \right)
\,,
\label{009}
\end{equation}
with the same survival probability,
because of CPT invariance
(we assume that $\nu_e$ and $\bar\nu_e$
have approximately the same energy spectrum).

The NuTeV collaboration reported that \cite{Zeller:2001hh}
\begin{equation}
\frac
{ N_{\nu_e}^{\mathrm{MC}} }
{ N^{\nu}_{\mathrm{L}} + N^{\nu}_{\mathrm{S}} }
\simeq
\frac
{ N_{\bar\nu_e}^{\mathrm{MC}} }
{ N^{\bar\nu}_{\mathrm{L}} + N^{\bar\nu}_{\mathrm{S}} }
\simeq
1.7 \times 10^{-2}
\,.
\label{010}
\end{equation}
In a first approximation we have
\begin{equation}
\frac{ N_{\nu_e}^{\mathrm{MC}} }{ N_{\nu_\mu}^{\mathrm{CC}} }
=
\frac
{ N_{\nu_e}^{\mathrm{MC}} }
{ N_{\nu_\mu}^{\mathrm{NC}} + N_{\nu_\mu}^{\mathrm{CC}} }
\,
\frac
{ N_{\nu_\mu}^{\mathrm{NC}} + N_{\nu_\mu}^{\mathrm{CC}} }
{ N_{\nu_\mu}^{\mathrm{CC}} }
\simeq
\frac
{ N_{\nu_e}^{\mathrm{MC}} }
{ N^{\nu}_{\mathrm{L}} + N^{\nu}_{\mathrm{S}} }
\left(
1 + R^{\nu}_{\mathrm{NuTeV}}
\right)
\,,
\label{011}
\end{equation}
and a similar relation for
$
N_{\bar\nu_e}^{\mathrm{MC}} / N_{\bar\nu_\mu}^{\mathrm{CC}}
$.
From Eq.~(\ref{010}) and the values
$R^{\nu}_{\mathrm{NuTeV}} \simeq R^{\bar\nu}_{\mathrm{NuTeV}} \simeq 0.4$
reported by the NuTeV collaboration \cite{Zeller:2001hh},
we obtain
\begin{equation}
\frac{ N_{\nu_e}^{\mathrm{MC}} }{ N_{\nu_\mu}^{\mathrm{CC}} }
\simeq
\frac{ N_{\bar\nu_e}^{\mathrm{MC}} }{ N_{\bar\nu_\mu}^{\mathrm{CC}} }
\simeq
2.4 \times 10^{-2}
\,.
\label{012}
\end{equation}

From Eqs.~(\ref{0061}), (\ref{008}), (\ref{009}), (\ref{012})
and the Paschos-Wolfenstein relation (\ref{003}),
the value
$\sin^2\theta_W^{\mathrm{NuTeV}}$
reported by the NuTeV collaboration is given by
\begin{equation}
\sin^2\theta_W^{\mathrm{NuTeV}}
=
\frac{1}{2}
-
\frac{ R^{\nu}_{\mathrm{NuTeV}} - r \, R^{\bar\nu}_{\mathrm{NuTeV}} }{ 1 - r }
=
\sin^2\theta_W
+
\frac{ N_{\nu_e}^{\mathrm{MC}} }{ N_{\nu_\mu}^{\mathrm{CC}} }
\,
P_{\nu_e\to\nu_s}
\,.
\label{013}
\end{equation}
Using the numerical values in Eqs.~(\ref{001}) and (\ref{012}),
we finally obtain
\begin{equation}
P_{\nu_e\to\nu_s}
\simeq
0.21 \pm 0.07
\,.
\label{014}
\end{equation}

The relatively large transition probability (\ref{014})
is compatible with the fact that reactor experiments
(see Ref.~\cite{Bemporad:2001qy})
did not measure a disappearance of electron antineutrinos
if the mass-squared difference
that drives the NuTeV oscillations is\footnote{
Our results may be compatible with those presented a long time ago in Ref.~\cite{Conforto:1990sp}.
}
\begin{equation}
\Delta{m}^2 \sim 10 - 100 \, \mathrm{eV}^2
\,,
\label{021}
\end{equation}
corresponding to an oscillation length in reactor experiments
\begin{equation}
L_{\mathrm{osc}}^{\mathrm{reactors}}
=
\frac{ 4 \pi \, E_{\mathrm{reactors}} }{ \Delta{m}^2 }
\sim
1 - 10 \, \mathrm{cm}
\,,
\label{022}
\end{equation}
where
$E_{\mathrm{reactors}} \sim 1 \, \mathrm{MeV}$.
Since the source-detector distance in reactor neutrino oscillation experiments
is larger than about 10 m,
the average survival probability of
electron antineutrinos
is a constant
and cannot be measured without assuming a precise knowledge of the initial $\bar\nu_e$ flux.

The oscillation length corresponding to a $\Delta{m}^2$ in the range (\ref{021})
for an energy of 100 GeV, typical of the NuTeV experiment,
is
\begin{equation}
L_{\mathrm{osc}}^{\mathrm{NuTeV}}
\sim
1 - 10 \, \mathrm{km}
\,,
\label{0221}
\end{equation}
which is appropriate for the
observation of transitions
with the NuTeV source-detector distance of 1.5 km.

Disappearance of electron neutrinos driven by  the $\Delta{m}^2$
given in Eq.~(\ref{021}) is particularly attractive for measurements
in the near future using the novel beta beam technique
proposed in Ref.~\cite{Zucchelli:2001gp},
with a source-detector distance of
the order of 10 m for a neutrino energy
$E \gtrsim 0.1 - 1 \, \mathrm{GeV}$.

These $\nu_e\to\nu_s$ transitions
could also be observed in other high precision
accelerator short-baseline neutrino oscillation experiments
like NOMAD \cite{Astier:2001yj},
with a source-detector distance of the order of 1 km
for a neutrino energy
$
E
\gtrsim
10 - 100 \, \mathrm{GeV}
$.

A high precision measurement could be done
in the long term at a
neutrino factory
(see Ref.~\cite{Albright:2000xi,Blondel:2000gj})
with a short baseline detector
located at a source-detector distance of 100 m
with a neutrino energy
$E \gtrsim 1 - 10 \, \mathrm{GeV}$
\cite{Bueno:2000jy,Adams:2001tv}.

The $\Delta{m}^2$ in Eq.~(\ref{021})
could be the same that drives the small $\bar\nu_\mu\to\bar\nu_e$
oscillations observed in the LSND experiment
\cite{Aguilar:2001ty}.
In this case,
$\nu_\mu\to\nu_e$
appearance
will be observed in the coming  MiniBooNE experiment
\cite{Church:1997ry}.
Simultaneous
$\nu_e\to\nu_s$ disappearance
and
$\bar\nu_\mu\to\bar\nu_e$
appearance
from neutrinos produced in
$\mu^+\to\bar\nu_\mu+e^++\nu_e$ decays\footnote{
Or
$\bar\nu_e\to\bar\nu_s$ disappearance
and
$\nu_\mu\to\nu_e$
appearance
from neutrinos produced in $\mu^-\to\nu_\mu+e^-+\bar\nu_e$ decays.
}
driven
by  the same $\Delta{m}^2$ given in Eq.~(\ref{021})
could be easily observed at a
neutrino factory with a short baseline detector
with electron charge discrimination capability
\cite{Bueno:2000jy}.
The signature of $\nu_e\to\nu_s$ oscillations will be a
depletion in the $e^-$ energy spectrum induced by $\nu_e$'s
and the presence of $\bar\nu_\mu\to\bar\nu_e$
oscillations will be characterized unequivocally by the
appearance of $e^+$ events.

$\nu_e\to\nu_s$
transitions driven by the high $\Delta{m}^2$ in Eq.~(\ref{021})
contribute with an energy-independent probability
to the disappearance of solar electron neutrinos
and must be taken into account in the analysis of solar neutrino data
(see \cite{BGG-review-98,Gonzalez-Garcia:2002dz}).
The energy dependence of the solar electron neutrino disappearance probability
observed in solar neutrino experiments
could be due to another very small squared-mass difference,
in the framework of four-neutrino mixing schemes
compatible also with the experimental evidence of atmospheric neutrino oscillations
(see \cite{BGG-review-98,Gonzalez-Garcia:2002dz,Neutrino_Unbound}).

The range (\ref{021})
for $\Delta{m}^2$ is compatible with the upper limit
$m_{\nu_e} \lesssim 3 \, \mathrm{eV}$
found in tritium decay experiments
(see Ref.~\cite{PDG}).
Indeed,
the averaged probability (\ref{014})
implies approximately
\begin{equation}
\sin^2 2 \vartheta \sim 0.4
\,,
\label{023}
\end{equation}
where $\vartheta$ is the mixing angle.
Assuming a hierarchy of neutrino masses,
the effective electron neutrino mass in tritium beta-decay experiments is
\begin{equation}
m_{\nu_e}
\simeq
\sin^2\vartheta \, \sqrt{\Delta{m}^2}
\sim
0.3 - 1 \, \mathrm{eV}
\,,
\label{024}
\end{equation}
that is compatible with the experimental upper bound.

If there is a hierarchy of
neutrino masses,
the effective neutrino mass in neutrinoless double-beta decay
is equal to $m_{\nu_e}$ in Eq.~(\ref{024}).
This value is compatible with the existing experimental upper limits
\cite{Klapdor-Kleingrothaus:2001yx,Aalseth:2002rf}
and curiously overlaps with the evidence in favor of
neutrinoless double-beta decay
claimed in Ref.~\cite{Klapdor-Kleingrothaus:2002ke}
(see, however, the discussion in Refs.~\cite{Aalseth-0202018,Feruglio-0201291}).

The large neutrino mass implied by the $\Delta{m}^2$ in Eq.~(\ref{021})
is compatible with the cosmological upper bound
on light neutrino masses
(about 10 eV),
but would have important effects
on the evolution of the universe
(see Ref.~\cite{Dolgov:2002wy}).

In conclusion,
we have propose an explanation of the NuTeV
in terms of $\nu_e\to\nu_s$ oscillations,
with the relatively large average transition probability
given in Eq.~(\ref{014})
and a large squared-mass difference given in Eq.~(\ref{021}).
These transitions are compatible with the results of
other neutrino experiments
and could be verified by future high-precision experiments.
Let us finally remark again that
our analysis has been approximate and based on the limited information
available in the paper of the NuTeV collaboration \cite{Zeller:2001hh}.
We hope that the NuTeV collaboration will consider seriously our suggestion
and perform a precise analysis of their data,
which could lead to a determination of the allowed ranges of
$\Delta{m}^2$ and $\sin^2 2 \vartheta$.


\begin{thebibliography}{10}

\bibitem{Zeller:2001hh}
NuTeV, G.~P. Zeller {\em et~al.},
\newblock (2001), hep-ex/0110059.

\bibitem{Davidson:2001ji}
S.~Davidson, S.~Forte, P.~Gambino, N.~Rius, and A.~Strumia,
\newblock (2001), hep-ph/0112302.

\bibitem{BGG-review-98}
S.~M. Bilenky, C.~Giunti, and W.~Grimus,
\newblock Prog. Part. Nucl. Phys. {\bf 43}, 1 (1999), hep-ph/9812360.

\bibitem{Gonzalez-Garcia:2002dz}
M.~C. Gonzalez-Garcia and Y.~Nir,
\newblock (2002), hep-ph/0202058.

\bibitem{Paschos:1973kj}
E.~A. Paschos and L.~Wolfenstein,
\newblock Phys. Rev. {\bf D7}, 91 (1973).

\bibitem{Bemporad:2001qy}
C.~Bemporad, G.~Gratta, and P.~Vogel,
\newblock (2001), hep-ph/0107277.

\bibitem{Conforto:1990sp}
G.~Conforto,
\newblock Nuovo Cim. {\bf A103}, 751 (1990).

\bibitem{Zucchelli:2001gp}
P.~Zucchelli,
\newblock (2001), hep-ex/0107006.

\bibitem{Astier:2001yj}
NOMAD, P.~Astier {\em et~al.},
\newblock Nucl. Phys. {\bf B611}, 3 (2001), hep-ex/0106102.

\bibitem{Albright:2000xi}
C.~Albright {\em et~al.},
\newblock (2000), arXiv:hep-ex/0008064.

\bibitem{Blondel:2000gj}
A.~Blondel {\em et~al.},
\newblock Nucl. Instrum. Meth. {\bf A451}, 102 (2000).

\bibitem{Bueno:2000jy}
A.~Bueno, M.~Campanelli, M.~Laveder, J.~Rico, and A.~Rubbia,
\newblock JHEP {\bf 06}, 032 (2001), hep-ph/0010308.

\bibitem{Adams:2001tv}
T.~Adams {\em et~al.},
\newblock (2001), hep-ph/0111030.

\bibitem{Aguilar:2001ty}
LSND, A.~Aguilar {\em et~al.},
\newblock Phys. Rev. {\bf D64}, 112007 (2001), hep-ex/0104049.

\bibitem{Church:1997ry}
BooNe, E.~Church {\em et~al.},
\newblock FERMILAB-P-0898.

\bibitem{Neutrino_Unbound}
C.~Giunti, M.~Laveder, and M.~Mezzetto,
\newblock (2002),
\newblock Neutrino Unbound {WWW} page:
  http://{\-}www.{\-}to.{\-}infn.{\-}it/{\-}\~{}giunti/NU.

\bibitem{PDG}
D.~E. Groom {\em et~al.},
\newblock Eur. Phys. J. {\bf C15}, 1 (2000),
\newblock WWW page: http://pdg.lbl.gov.

\bibitem{Klapdor-Kleingrothaus:2001yx}
H.~V. Klapdor-Kleingrothaus {\em et~al.},
\newblock Eur. Phys. J. {\bf A12}, 147 (2001).

\bibitem{Aalseth:2002rf}
C.~E. Aalseth {\em et~al.},
\newblock (2002), hep-ex/0202026.

\bibitem{Klapdor-Kleingrothaus:2002ke}
H.~V. Klapdor-Kleingrothaus, A.~Dietz, H.~L. Harney, and I.~V. Krivosheina,
\newblock Mod. Phys. Lett. {\bf A16}, 2409 (2002), hep-ph/0201231.

\bibitem{Aalseth-0202018}
C.~E. Aalseth {\em et~al.},
\newblock (2002), hep-ex/0202018.

\bibitem{Feruglio-0201291}
F.~Feruglio, A.~Strumia, and F.~Vissani,
\newblock (2002), hep-ph/0201291.

\bibitem{Dolgov:2002wy}
A.~D. Dolgov,
\newblock (2002), hep-ph/0202122.

\end{thebibliography}
\end{document}